\documentclass[letter,twocolumn]{jpsj3}
\usepackage{bm,color}

\title{Sub-gap in the Edge States of 2-D Chiral Superconductor
with Rough Surface}

\author{Yasushi \textsc{Nagato}, 
Seiji \textsc{Higashitani} and Katsuhiko \textsc{Nagai} }

\inst{Graduate School of Integrated Arts and Sciences, Hiroshima
University,
Kagamiyama 1-7-1, Higashihiroshima 739-8521, Japan
}

\abst{
We discuss the rough surface effects on a two-dimensional chiral $k_x+ik_y$
 superconductor. The atomic scale roughness at the surface is
considered using the random $S$ matrix model. The roughness effects
on the self-consistent order parameter, the surface mass current and
the surface density of states are studied using the quasi-classical
theory. We find that the surface mass current is suppressed by the
surface roughness. The surface density of states shows a quite similar
behavior to that of superfluid ${}^3$He B phase. When the surface is
 specular,
the surface Andreev bound states form a band which fills the bulk
energy gap $\Delta_{\rm bulk}$. When the surface becomes diffusive,
there occurs a sharp upper edge of the surface bound states band and there
opens a sub-gap between the edge and the bulk energy gap. We show
that this sub-gap  is induced by the repulsion between the surface
bound states and the propagating Bogoliubov quasi-particles
through the second order process of roughness. 
}

\kword{2-D Chiral Superconductor, Edge States, Surface Demsity of
States,
Superfluid ${}^3$He-B}

\usepackage{bm}
\begin{document}
\maketitle

The 2-dimensional chiral $k_x+ik_y$ state is known as a model system
for Sr${}_2$RuO$_4$ superconducting state.\cite{Maeno} When there is a surface
along the $y$ axis, there occur gapless surface Andreev bound states
as in other $p$-wave\cite{BZ,Hara} and $d$-wave\cite{Hu} pairing systems,
because the $k_x$ component of the order parameter changes its sign
under the surface reflection. The surface bound states in chiral
system are known to carry spontaneous mass flow along the surface.\cite{MS,SR}
Recently, surface bound states are
recognized as edge states which reflect the topological nature
of the bulk pairing state. A lot of attention has been paid to the 
surface Andreev bound states from this
aspect.\cite{SR,volovikc,Schnyder,Roy,Kitaev,Qi,Chung,Volovik,Nagato09,
MurakawaJPSJ}

In this paper, we consider the effects of atomic scale surface roughness
on the chiral $k_x+ik_y$ state. We use the quasi-classical
theory\cite{Nagato1996,Nagato} developed for the study of $p$-wave Fermi
superfluids. We calculate the self-consistent order parameter, the
surface mass current and the surface density of states. The surface
density of states shows quite a similar behavior to that in the
3-dimensional
 BW state.\cite{Nagato} Existence of the order parameter component
parallel
to the surface disperses the surface bound state energy. The bound
states,
therefore, form a band below the bulk energy gap $\Delta_{\rm bulk}$.
When the surface is specular, the  band completely 
fills the bulk energy gap. When the surface becomes diffusive,
however,
there occurs a sharp upper edge of the band and there
opens a sub-gap between the edge and the bulk energy gap. The band edge
energy $\Delta^*$ increases as the roughness is reduced.
Similar sub-gap has
been
known in the BW state.\cite{Zhang, Nagato, Vorontsov}
The sub-gap was first 
reported by Zhang\cite{Zhang} who treated the surface roughness using the thin
dirty layer model. He suggested that the sub-gap is due to the suppression
of the parallel component of the order parameter by the roughness.
However, Nagato et al.\cite{Nagato} found that this sub-gap also occurs when the
order parameter is assumed to be spatially constant. Although the existence of
the
sub-gap played a decisive role in the interpretation of the 
transverse acoustic impedance of the B phase of
superfluid ${}^3$He,\cite{Aoki,MurakawaPRL,Nagato2007,Nagai}  
the origin of the sub-gap has been a
puzzle for a long time.

We consider a two-dimensional $k_x+ik_y$ superconductor
which fills the $x>0$ domain. The surface extends along the $y$ axis.
Since the order parameter is suppressed near the surface, the order
parameter
 will take a form
\begin{equation}
 \Delta(\hat{\bm{k}},x)=\Delta_{\perp}(x)\hat{k}_x+i\Delta_\parallel(x)\hat{k}_y
=\Delta_{\perp}(x)\cos\phi+i\Delta_\parallel(x)\sin\phi,
\end{equation}
where $\phi$ is the angle between the Fermi momentum and the $x$ axis.

We investigate the effects by
surface roughness of atomic scale using random S-matrix 
model.\cite{Nagato1996,Nagato,Nagai}
The surface is characterized by an $S$-matrix for the quasi-particles
at the Fermi level in the normal state
\begin{equation}
S_{k_y,q_y}=-\left(\frac{1-i\eta}{1+i\eta}\right)_{k_yq_y},
\end{equation}
where $k_y (q_y)$ is the $y$ component of the incident (scattered)
Fermi momentum and $\eta$ is a Hermite matrix that specifies the
surface roughness. We assume that $\eta$ is a random Hermite matrix
which obeys $\overline{\eta_{k_y q_y}}=0$ and 
$\overline{\eta_{k_y q_y}^*\eta_{{k'}_y{q'}_y}}=2W/(\sum_{q_y}1)
\delta_{k_y-q_y,{k'}_y-{q'}_y}$ with $W$ a parameter
that specifies the roughness of the surface.   One can show
that $W=1$ corresponds to the diffusive surface boundary condition and
$W=0$ corresponds to the specular surface boundary 
condition.\cite{Nagato1996,Nagato}  

Taking into account the surface roughness within the self-consistent Born
approximation, we  obtain the quasi-classical Green's
function at the surface.\cite{Nagato,Nagai}
\begin{align}
 G_{\pm}(k_y,0)&=G_S+(G_S\pm i)\dfrac{1}{G_S^{-1}-\Sigma}(G_S \mp i),
\label{green}\\
 \Sigma&= 2W \left<\dfrac{1}{G_S^{-1}-\Sigma}\right>.\label{self}
\end{align}
Here $G_{+}(k_y,0)$ and $G_{-}(k_y,0)$ 
are the quasi-classical Green's function
for the Fermi momentum $(k_x>0,k_y)$ and $(k_x<0, k_y)$, respectively.
$G_S$ is the quasi-classical
Green's function at $x=0$ for the specular surface and
$\Sigma$ is the surface self energy which is induced by the roughness.
The angle average in the two-dimensional system is
\begin{align}
 <\cdots>&=\dfrac{\int dk_y \cdots}{\int dk_y 1}=\frac{1}{2}\int_{-\pi/2}^{\pi/2} d\phi\cos\phi\  \cdots .
\end{align}
The quasi-classical Green's function $G_\pm(k_y,x)$ at finite $x$ is
calculated by evolution operator technique.\cite{Nagato1996,Nagato}

\begin{figure}
\includegraphics[width=8cm]{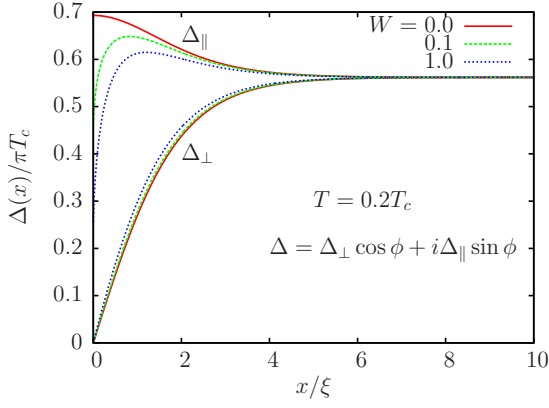}
\caption{(color online) Self-consistent order parameter of $k_x+ik_y$ state is plotted
against the distance from the surface scaled by the coherence length
$\xi=v_F/\pi T_c$ for some typical values of the roughness parameter $W$.
}
\label{OD}
\end{figure}

Using the Matsubara Green's function, we can calculate the
self-consistent
order parameter and the edge mass current. In Fig.~\ref{OD}, we show the
self-consistent order parameter at $T=0.2T_c$. 
Since the bulk energy gap 
is isotropic in the two-dimensional $k_x+ik_y$ state, the order
parameter
shows a quite similar profile to that of the three-dimensional BW state. The
perpendicular component $\Delta_{\perp}(x)$ is suppressed near the surface.
In case of the specular surface,\cite{MS} the parallel component $\Delta_\parallel(x)$
is
enhanced near the surface such that compensates the loss of the
condensation
energy caused by the suppression of the perpendicular component $\Delta_{\perp}(x)$.
In case of the diffusive surface, $\Delta_\parallel(x)$ is also
suppressed by
the incoherent phase mixing during
the reflection processes.\cite{AK}

\begin{figure}
 \includegraphics[width=8cm]{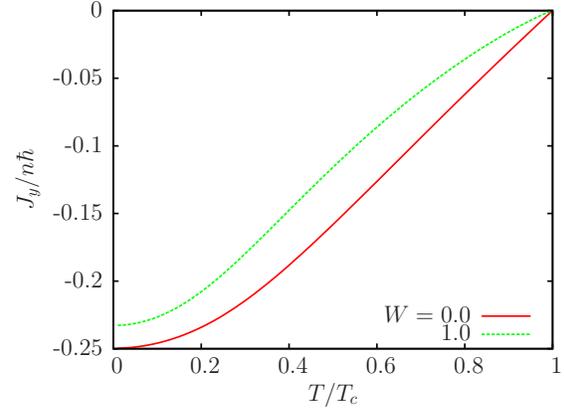}
\caption{ (color online) Temperature dependence of the total current $J_y$ along the $y$
 axis. 
Solid curve is for the specular surface $(W=0)$ and the dotted curve is
 for the diffusive surface $(W=1)$. 
}
\label{totalcurrent}
\end{figure}

Once the order parameter is determined,
the surface mass current along the $y$ axis 
can be calculated from the diagonal element of
the quasi-classical Green's function. In Fig.~\ref{totalcurrent}, we
show
the total surface mass current $J_y$, current density integrated over $x$,
as a function of temperature. In case of the specular surface $(W=0)$,
the total current tends to $J_y=-n\hbar/4$ as $T\rightarrow 0$K.\cite{SR}
Here, $n$ is the total number density. When the surface is diffusive $(W=1)$, the total current
is
definitely suppressed. The suppression of the mass current by surface roughness was discussed
by Ashby and Kallin\cite{AK} using GL theory.

\begin{figure}[h]
\includegraphics[width=8cm]{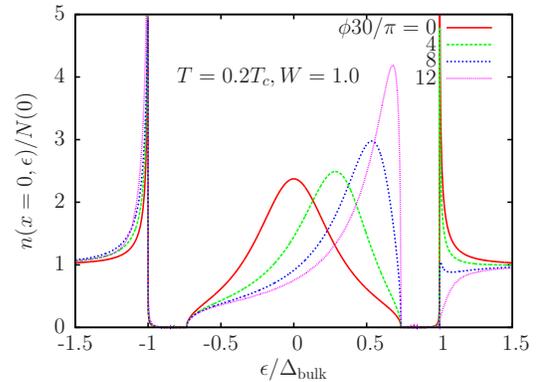}
\caption{(color online) Angle resolved surface density of states in case of
 the diffusive
surface ($W=1$). Incident angles are $\phi=0,\frac{4}{30}\pi,\frac{8}{30}\pi$ and $\frac{12}{30}\pi$.
}
\label{dens_d}
\end{figure}

The surface density of states can be calculated from the quasi-classical
Green's function 
with
real frequency $\epsilon$. We show the angle resolved density of states
in Fig.~\ref{dens_d}
for the diffusive surface ($W=1$).
In case of the specular surface, the surface density of states shows
a delta function peak that corresponds to the surface Andreev bound
state.
The peak position is roughly equal to $\Delta_\parallel(0)\sin\phi$. When
integrated over the angle, therefore, the bulk energy gap below
$\Delta_{\rm bulk}$ is filled by the bound states. In case of the
diffusive surface, the bound state peak is broadened and is shifted
towards the lower energy. Moreover, there appears a sharp upper energy
edge $\Delta^*$ common to all the incident angles, which leads to a
sub-gap between $\Delta^*$ and $\Delta_{\rm bulk}$.

To examine the origin of the sub-gap, we consider 
Green's function at the surface given by Eqs.~(\ref{green}) and
(\ref{self}).
To discuss the density of states below the
bulk energy gap $\Delta_{\rm bulk}$, we consider real frequency
$|\epsilon| < \Delta_{\rm bulk}$.
Let us first consider $G_S$ which is given by\cite{Nagato,Nagai}
\begin{align}
 G_S&=\dfrac{1}{1-D^2}\begin{pmatrix}
i(1+D^2) & -2D\\
-2D & -i(1+D^2)\end{pmatrix},\label{gs}
\end{align}
where $D=D(0,\epsilon,k_y)$ is a solution at $x=0$ of the Ricatti equation
\begin{align}
 v_F\cos\phi\dfrac{\partial}{\partial x}D&=-2i\epsilon D+\Delta(\hat{k},x)D^2-\Delta^*(\hat{k},x)\label{ricatti}
\end{align}
with the boundary condition at the bulk infinity
\begin{align}
D(\infty,\epsilon,k_y)&=\dfrac{i\Delta^*(\hat{k},\infty)}{\epsilon+\sqrt{\epsilon^2-|\Delta_{\rm
 bulk}|^2}}=e^{i(\alpha-\phi)},\label{boundary}
\end{align}
where we have defined $ \alpha=\sin^{-1}(\epsilon/\Delta_{\rm bulk})$.
For the energy $|\epsilon|<\Delta_{\rm bulk}$,
it can be shown from Eq.~(\ref{ricatti}) that $|D(x,\epsilon,k_y)|$ is always unity,
therefore we may write
\begin{align}
D(x,\epsilon,k_y)=e^{i\theta(x,\epsilon,k_y)}\label{ddd}
\end{align}
with $\theta$ the real function.
Solving Eq.~(\ref{ricatti}) for a given $k_y$, 
we find an $\epsilon$ that satisfies
\begin{align}
 D(0,\epsilon,k_y)&=1.
\end{align}
This energy is the surface bound state energy for the specular surface 
because the Green's function
$G_S$ has a pole at that energy. 
It is worth noting that $D$ is related to the Nambu amplitude
$(u(x),v(x))$
of the state with energy $\epsilon$. The ratio of the hole
component
$v$ to the particle component $u$ is given by $v/u=(-i)D$.
It follows that the hole-particle ratio $v/u$ at the surface is 
equal to $-i$  for
all the surface bound states.

When we assume that the order parameters
are constant, i.e., $\Delta_{\perp}(x)=\Delta_\parallel(x)=\Delta_{\rm bulk}$, $D$ is also a
constant given by Eq.~(\ref{boundary}). The bound states have a linear
dispersion relation
$
 \epsilon=\Delta_{\rm bulk}\sin\phi=\Delta_{\rm bulk}\hat{k}_y
$ and can be regarded as Majorana-Weyl Fermions.\cite{SR}

Now we consider the surface self energy. From Eqs.~(\ref{self}) and (\ref{gs}),
we may parametrize the self energy in a form
\begin{align}
 \Sigma&=\begin{pmatrix}i s_3 & s_1 \\ s_1 & -is_3\end{pmatrix}.
\end{align}
It is convenient to introduce  projection operators 
$ P_\pm=\frac{1}{2}(1\pm \rho_2)$
with $\rho_2$  a Pauli matrix in particle-hole space.
Then we can write
\begin{align}
 \Sigma&=i\rho_3\left(P_+(s_3+s_1)+P_-(s_3-s_1)\right)\label{proj1},\\
 G_s&=i\rho_3\left(P_+\frac{1-D}{1+D}+P_-\frac{1+D}{1-D}\right).\label{proj2}
\end{align} 
It is obvious from Eq.~(\ref{proj2}) that $P_+$ projects out the surface bound states.
From Eqs.~(\ref{self}), (\ref{proj1}) and (\ref{proj2}), we find that
\begin{align}
 s_3+s_1&=2W\left<
\dfrac{1-D}{\left(1+(s_3-s_1)\right)+\left(1-(s_3-s_1)\right)D}
\right>,\label{selfp}\\
s_3-s_1&=2W\left<
\dfrac{1+D}{\left(1+(s_3+s_1)\right)-\left(1-(s_3+s_1)\right)D}
\right>.\label{selfm}
\end{align}

The density of states is given by the imaginary part of the
diagonal element of the
quasi-classical Green's function  given by Eqs.~(\ref{green}), (\ref{proj1})
and (\ref{proj2}).\cite{Nagato}
\begin{align}
G^{11}_+ =&G^{11}_- \nonumber\\
 =&\frac{i}{2}
\left(
\dfrac{\left(1+(s_3+s_1)\right)+\left(1-(s_3+s_1)\right)D}
{\left(1+(s_3+s_1)\right)-\left(1-(s_3+s_1)\right)D}\right.\nonumber\\
 &+\left.
\dfrac{\left(1+(s_3-s_1)\right)-\left(1-(s_3-s_1)\right)D}
{\left(1+(s_3-s_1)\right)+\left(1-(s_3-s_1)\right)D}
\right).\label{diagonal}
\end{align}
When the energy is in the range $|\epsilon|<\Delta_{\rm bulk}$,
$D$ is given from Eq.~(\ref{ddd}) by a form
$e^{i\theta(0,\epsilon,k_y)}$. 
It follows that
if both $s_3$ and $s_1$ are pure imaginary, the diagonal element of the
Green's function is real, namely there is no density of states. At the
sub-gap energies, therefore, $s_3, s_1$ are expected to take pure
imaginary values. At first sight, both Eqs.~(\ref{selfp}) and (\ref{selfm})
have pure imaginary solutions. Both the equations are 
invariant under the
complex conjugate transformation because $D=e^{i\theta}$. The real part
emerges when there appears a pole along the angle integral in 
Eqs.~(\ref{selfp}) and (\ref{selfm}). 

From now on, for simplicity, we consider a case where the order parameters are constant
and the roughness parameter
$W$ is small. When the order parameter is constant, $D=e^{i(\alpha-\phi)}$;
therefore, the bound state
energy for the specular surface is given from $D=1$ by
$ \alpha=\phi\  (\epsilon=\Delta_{\rm bulk}\sin\phi)$. 
Within the lowest order correction with respect to $W$,
$s_3+s_1$ remains pure imaginary because the bound states are projected
out in Eq.~(\ref{selfp}).
\begin{align}
 s_3+s_1&=2W\left<(-i){\rm Tr}(P_+\rho_3G_s)\right>=2W\left<\frac{1-D}{1+D}\right>\label{Born}\\
 &=iW\left(\cos\alpha\ln\left|\dfrac{1+\sin\alpha}{1-\sin\alpha}\right|
-\pi \sin\alpha\right)\equiv\frac{i}{2}f(\alpha).\label{eqf}
\end{align}
On the other hand, $s_3-s_1$ acquires real part because
Eq.~(\ref{selfm})
has a pole of the surface bound state when $s_3+s_1$ is neglected. Let us consider the next order
correction by $W$ to the possible pole of Eq.~(\ref{selfm})
\begin{align}
 D&=\dfrac{1+(s_3+s_1)}{1-(s_3+s_1)}.\label{pole1}
\end{align}
When $W$ is small, the possible pole will occur near $D\sim 1+i(\alpha-\phi)$.
Expanding both sides of Eq.~(\ref{pole1}) in terms of small quantities, 
we obtain
\begin{equation}
 \alpha-\phi=(-2i)(s_3+s_1)= f(\alpha). \label{eqimportant}
\end{equation}
 
\begin{figure}[h]
\begin{center}
 \includegraphics[width=6cm]{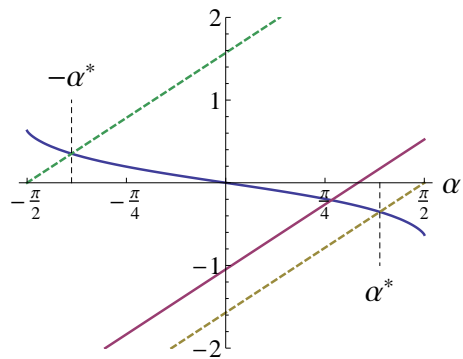}
\end{center}
\caption{(color online) $\alpha-\phi\ (\phi=-\pi/2, \pi/3, \pi/2)$ and $f(\alpha)$ with $W=0.1$ are plotted.}
\label{BW}
\end{figure}
\begin{figure}[h]
 \begin{center}
  \includegraphics[width=8cm]{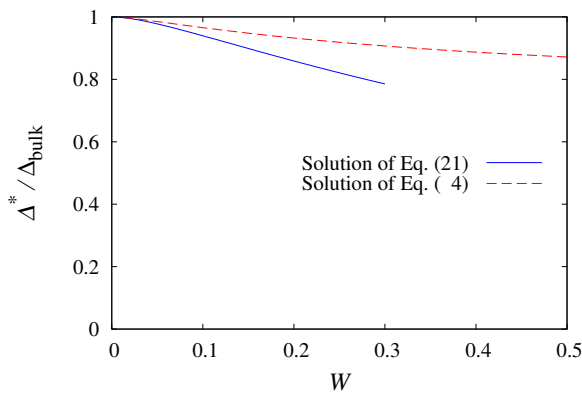}
 \end{center}
\caption{(color online) $\Delta^*$ vs $W$. Order parameters are
assumed to be spatially constant. Dashed curve is a result of self-consistent
solution of Eq.~(\ref{self}). Solid curve is a result of Eq.~(\ref{astar}).}
\label{aastar}
\end{figure}

We plot the both hand sides of Eq.~(\ref{eqimportant})
in Fig. \ref{BW} as functions of $\alpha=\sin^{-1}(\epsilon/\Delta_{\rm bulk})$.
Since $f(\alpha)$ is a decreasing odd function of $\alpha$,
there is a solution of Eq.~(\ref{eqimportant}) for any $-\pi/2 < \phi <
\pi/2$.
But, when we define $\alpha^*$ at which the straight line $\alpha-\pi/2$ and $f(\alpha)$ crosses;
\begin{align}
 \alpha^*-\pi/2&=f(\alpha^*),\label{astar}
\end{align}
we find that
there is no solution of Eq.~(\ref{eqimportant})
for $\alpha$ in the range $\alpha^*<|\alpha|<\pi/2$. It means that
Eq.(\ref{selfm}) has no pole and $s_3-s_1$ remains pure imaginary in that
energy range.
As a result, for all the incident angles there occurs a common 
sub-gap in the energy range
$\Delta^*=\Delta_{\rm bulk}\sin\alpha^* < |\epsilon| < \Delta_{\rm
bulk}$, as seen in Fig.~\ref{dens_d}.
Solving Eq.~(\ref{astar}) with respect to $\alpha^*$, we obtain 
$\Delta^*=\Delta_{\rm bulk}\sin\alpha^*$ as a function of $W$.
The result is plotted in Fig.~\ref{aastar} together
with the self-consistent solution of Eq.~(\ref{self}).

The origin of the sub-gap is interpreted in a following way.
Since Eq.~(\ref{pole1}) is also a possible pole of the
diagonal element $G_{\pm}^{11}$ of the Green's function 
(see Eq.~(\ref{diagonal})),
Eq.~(\ref{eqimportant}) is interpreted to be an equation to determine the
energy of the bound state with finite $W$, although we have used it to
find out the energy range without solution.
Equation (\ref{eqimportant}) with (\ref{Born}) has a similar form to the denominator
of the usual Green's function for the impurity problem within the Born
approximation, 
therefore it corresponds to the
Brillouin-Wigner perturbation formula
\begin{align}
 \epsilon-\epsilon_n^{(0)}&=\sum_m\dfrac{|V_{nm}|^2}{\epsilon-\epsilon_m^{(0)}},
\label{Brillouin}
\end{align}
where $\epsilon_n^{(0)}$ corresponds to $\Delta_{\rm bulk}\sin\phi$ and 
$|V_{nm}|^2$ to $W$.
Since the bound states are projected out in Eqs.~(\ref{Born}) and 
(\ref{eqimportant}), the
intermediate states are the propagating Bogoliubov quasi-particle states with energy
$|\epsilon|>\Delta_{\rm bulk}$. 
The right hand side of Eq.~(\ref{Brillouin}) becomes a decreasing odd
function of $\epsilon$ and reproduces the $\alpha$ dependence of $f(\alpha)$.
The sub-gap comes out, thus, as a result of the 
repulsion between the bound state and the propagating states through the
second order process.  This scenario
does not change in case of the self-consistent order parameter, although
 Eq.~(\ref{eqimportant}) should be calculated numerically.
The sub-gap in superfluid ${}^3$He-B can be explained in a similar manner.

It is of interest if the sub-gap which has been observed in 
the B phase of superfluid ${}^3$He can be also observed in
Sr$_2$RuO$_4$,
for example by tunneling experiment\cite{Laube, Kashiwaya}.
For comparison with experiment, the effects by finite transmittance 
of the rough interface should be examined.\cite{Nagatonagai}
Such a study shall be reported elsewhere.


We thank Y. Okuda for useful comments on the manuscript.
This work is supported in part by Grant-in-Aid for Scientific Research
(No. 21540365) from MEXT of Japan. One of the authors (S. H.) is also
supported by the ``Topological Quantum Phenomena''(No. 22103003) KAKENHI
on Innovative Areas from MEXT of Japan.


\end{document}